\documentstyle[aps,prl,epsf]{revtex}

\begin{document}



\draft
\twocolumn[\hsize\textwidth\columnwidth\hsize\csname
@twocolumnfalse\endcsname
\title{\bf Vortex avalanches and self organized criticality in
superconducting niobium}
\author{E. Altshuler$^{1,2}$, T.H. Johansen$^{3,2}$,\\ Y.
Paltiel$^4$,Peng Jin$^{2}$, K.E. Bassler$^{5}$, O. Ramos$^{1}$,\\
G.F. Reiter $^{5}$, E. Zeldov$^{4}$ and C.W. Chu$^{2,6,7}$}
\address{$^1$Superconductivity Laboratory\\ IMRE-Physics
Faculty, University of Havana, 10400 Havana, Cuba\\
$^2$Texas Center for Superconductivity, University of Houston,
Houston, TX 77204-5002, U.S.A.\\ $^3$Department of Physics,
University of Oslo, POB 1048 Blindern, N-0316 Oslo, Norway\\
$~^4$Condensed Matter Physics Department, Weizmann
Institute of Science, Rehovot 76100, Israel.\\ 
$~^5$Department of Physics, University of Houston, Houston, TX
77204-5005, U.S.A. \\ $~^6$Lawrence Berkeley National
Laboratory, Berkeley, CA 94720, U.S.A. \\ $~^6$Hong Kong
University of Science and Technology, Clear Water Bay, Kowloon,
Hong Kong.}
\date{\today}
\maketitle

\begin{abstract}
In 1993 Tang proposed \cite{Tang-1993} that vortex avalanches
should produce a self organized critical state in superconductors,
but conclusive evidence for this has heretofore been lacking. In
the present paper, we report extensive micro-Hall probe data from
the vortex dynamics in superconducting niobium, where a broad
distribution of avalanche sizes scaling as a power-law for more
than two decades is found. The measurements are combined with
magneto-optical imaging, and show that over a widely varying
magnetic landscape the scaling behaviour does not change, hence
establishing that the dynamics of superconducting vortices is a
SOC phenomenon.
\end{abstract}
]
\narrowtext

The concept of self-organized criticality (SOC), first proposed by
Bak, Tang and Weisenfeld \cite{BTW-1987}, and then applied to many
scenarios exhibiting avalanche dynamics
\cite{Bak-1996,Jensen-1998} refers to the tendency of a system to
endogenously self-organize into a critical state. An idealized
sandpile is the prototypical example of a SOC system. As grains of
sand are slowly added to the pile, it ``organizes" itself by tuning
the slope to a "critical" value, resulting in a marginally stable
stationary state. Avalanches are scale invariant in a SOC system,
and the distribution $P(s)$ of their size $s$ follows a power law
$s^{-\tau}$, where $\tau$ is a critical exponent. The value of
$\tau$ is said to be universal in the sense that it should be
robust to minor changes in the system.

Granularity, although of a different kind, is an inherent property
also of type-II superconductors. When an external magnetic field
is raised above a certain critical value, the material becomes a
host for quantized magnetic vortices, which nucleate at the edges
and are driven inwards by their mutual repulsion. In most
materials this motion is impeded by microscopic defects acting as
``pinning sites", with the result that the vortices pile up with a
characteristic density gradient, producing the so called ``critical
state" \cite{Bean-1964}. DeGennes first suggested as early as 1966
that avalanche dynamics could be also present in the
superconducting critical state \cite{DeGennes-1966}.

The possible development of SOC in type II superconductors plus
the fact that the motion of vortices is overdamped, making them an
ideal system in which to look for SOC behavior, has revived the
interest in vortex avalanches in the last years, starting with the
work of Field and coworkers \cite{Field-1995}. In those
experiments, and in similar ones performed by Heiden and Rochlin
\cite{Heiden-1968} vortex avalanches were detected by means of
large pick up coils in the center of Pb-In and Nb-Ti cylinders.
However, with this experimental setup only the flux that leaves
the superconductor is measured, and the amount of flux involved in
each avalanche event can not be directly measured. Direct
observation of avalanche dynamics with single vortex resolution is
today possible using micro-Hall probes. With this technique
Seidler et al. \cite{Seidler-1993} found that avalanche-like
behavior occurs at high fields and very low temperatures in
YBaCuO crystals. The avalanche statistics, later worked out in
\cite{Zieve-1996}, was quite limited being based on only a few
hundred events. Nowak et al. \cite{Nowak-1997} studied niobium
thin films where the statistics was improved, although avalanche
sizes below 50 vortices were absent. As pointed out by the
authors, such films exhibit a thermo-magnetic instability, or
catastrophic flux jumps, which may camouflage other basic
dynamical features. This is consistent with the results by Esquinazi
et al in thin Nb films \cite{Esquinazi et al-1999}. Recently, much
thicker niobium samples were investigated by Behnia et al.
\cite{Behnia-2000}, who found only very small avalanches under the
Hall probe area, leaving an open question whether the distribution
is a power-law or stretched exponential. Until now no conclusive
evidence has existed that the dynamics of magnetic vortices in
superconductors is a genuine realization of SOC.

Local avalanche measurements are largely improved by first mapping
out the global magnetic landscape in which the Hall probes are
placed. Therefore, we  made magneto-optical  (MO) imaging  of the
flux penetration in our superconducting sample, a Nb foil of
dimensions $1.5 \times 1.5 \times 0.25 mm^3$, and carefully
superimposed such images onto a picture taken afterwards of the
Hall probe arrangement attached to the large face of the sample.
The MO pictures pictures are obtained using a Faraday-active
ferrite garnet film mounted on top of the superconductor, which is
viewed through crossed polarizers in a microscope. Fig. 1 shows
the MO image recorded at $4.8 K$ with a magnetic field of $400 Oe$
applied perpendicular to the foil. The flux penetration is seen to
advance in finger-like structures producing a ridged landscape,
where each ridge has the characteristics of the critical-state
with an ``inverted V" flux density profile (the MO investigation
also showed that the finger structure grows smoothly as the field
is increased, and should not be confused with the dendritic
structures abruptly appearing during thermal runaways in thin
films \cite{Duran-1995,Leiderer-1993,Johansen-2002}). An array of
11 Hall-sensors (white dots) was mounted along the large central
ridge in the landscape, which we regard an optimum location for
avalanche observations.

Shown in Fig. 2 are typical data obtained from one of the Hall
sensors while a magnetic field perpendicular to the Nb foil was
ramped from 0 to $3.5 kOe$ at the rate of $1 Oe/s \pm 0.05 Oe/s$
using a superconducting magnet. The resolution of the micro-Hall
probe, which is made from a GaAl/AlGaAs heterostructure, is better
than $0.21 G$, equivalent to one flux quantum  $(\Phi_o = 2.1
\times 10^{-15} Tm^2)$ under the probe area.We quantify the Hall
signal by the number of vortices (flux quanta) located under the
$10 \times 10 mm^2$ sensor area. Although the overall curve
appears smooth, the magnified views in the insets clearly show
that the amount of flux in a given area varies in distinct steps.
This is direct evidence that the vortex penetration evolves by an
avalanche-type of dynamics. The individual steps, which we define
as avalanche events, have a vertical size ranging from fractions
of a flux quantum to more than a hundred vortices. Near $400 Oe$
the main curve makes an upturn, which indicates when the
propagating flux front reached the probe position. Above $1 kOe$
the Hall signal grows linearly, corresponding to the Nb foil being
fully penetrated. The whole curve follows the prediction of the
critical state model for perpendicular geometries
\cite{Zeldov-1994}.

The signals recorded from all the 11 probes were subjected to
further statistical analysis. A histogram of the avalanche sizes,
taken as the number of magnetic flux quanta in a distinct step, is
shown in Fig.3. The data include approximately 200 000 avalanche
events observed during several field sweeps up to $3.5 kOe$
repeated under identical external conditions. The number of
avalanches, $P(s)$, versus their size is seen to behave as a
power-law over two and a half orders of magnitude in s, with a
critical exponent of $\tau = 3.0 \pm 0.2$. Note that our total
number of avalanche events is about 100 times larger than in
previous studies using micro Hall probes
\cite{Seidler-1993,Zieve-1996,Nowak-1997,Behnia-2000}, and our
distribution function spans more than 6 decades, which is
essential for establishing a power-law distribution in this type
of systems.

To address the question of robustness, we performed avalanche
statistics at a variety of locations over the sample area. From
Fig. 4, where the different probe locations are indicated, one
sees e.g. that while all probes in arrangement II are parallel to
the same central finger as arrangement I, also avalanches coming
from the neighboring ``hillside" will contribute. All probes in
arrangement III are located parallel to two "competing" fingers,
which, due to some degree of anisotropy in the sample, are thicker
than those developing from the upper sample edge. A very different
location is the one corresponding to Hall probes 1, 2, 3 and 4 on
arrangement IV, and probes 5, 6, 7 and 8 on arrangement V
(counting from the left), which was also investigated. A
remarkable result is that the avalanche size statistics performed
on each of the above mentioned groups of probes gave quite good
power laws within nearly one and a half decades on the horizontal
axis, and with critical exponent $\tau$ consistent with the value
reported above.

Although the avalanche distribution is universal in SOC models
(see, for example, ref. 18), the slope of the flux density is not.
Instead, the slope of the flux density varies with the system
parameters such as the local density, or strength, of pins. The
topography we observe in Fig. 1 is reproducible from run to run,
and hence the deviations we observe from a uniform penetration are
due to quenched disorder. Nevertheless, the value of the critical
exponent $\tau$ we measure is the same at all points in the
sample, even though the local magnetic topography in the region of
the probe varies from point to point. The fact that the exponents
found in the present work are all essentially the same is
characteristic of the universality expected for SOC. This
robustness, together with the power law distribution of avalanche
sizes over two and a half decades provides convincing evidence
that SOC is a correct description of the penetration of flux into
a type II superconductor in the presence of strong disorder.

We appreciate the discussions with, P. Bak, A.J. Batista-Leyva, H.
Jaeger, B. Lorenz, R. Mulet, M. Paczuski, D. V. Shantsev, A.
Malthe-S{\o}renssen and Y. Xue. The authors thank the support by
the World Laboratory Center for Pan-American Collaboration in
Science and Technology, the Research Council of Norway, the
National Science Foundation, the Israel Science Foundation -
Center of Excellence Program, the ``Alma Mater" grants program
(University of Havana), and the Department of Energy (U.S.A.).

\newpage


\subsection*{Figure Captions}

\vspace{0.2truein}

Figure 1 Magnetic landscape of the Nb superconductor. The
magneto-optical image (top) shows how the flux penetrates into one
half of a $1.5 \times 1.5 \times 0.25 mm^3$ Nb foil at a field of
400 Oe applied perpendicular to the sample at $T = 4.8 K$. The
image brightness represents the local density of the magnetic
vortices. The 3D-plot (bottom) of the penetration pattern shows
that the magnetic landscape consists of several ridges with smooth
slopes, which rise up from the flat Meissner state (flux free)
area. Included in the figure is the location of a Hall probe array
where each of the 11 probes detects the flux under an area of $10
\times 10 mm^2$. As the applied magnetic field increases all the
ridges grow gradually on a macroscopic scale, quite similarly to
the way sandpiles increase in size when grains are added.

\vspace{0.2truein}

Figure 2 Vortex avalanches seen by a Hall probe during field
increase. The main curve was obtained from probe number 4 from the
sample edge in Fig. 1, and contains more than 40,000 data points.
The insets show zooms in two different field windows revealing
distinct steps as a manifestation of avalanche dynamics. The
experiment was performed at $T = 4.8 K$.

\vspace{0.2truein}

Figure 3  Size distribution of avalanches. The statistics is based
on nearly 200 000 avalanche events measured in the Hall probes
located as shown in Fig. 1. Steps less than $0.5 \Phi_o$ were
excluded from the counting of events. The avalanche sizes were
exponentially binned, giving equi-distant points in the log-log
plot. The data are fitted by the straight line $P(s)\sim
s^{-\tau}$, with $\tau = 3.0 \pm 0.2$.

\vspace{0.2truein}

Figure 4 Magneto-optical image of the sample with an overlay
showing 5 different locations of the Hall probe array used for
supplementary measurements. Location (I) is identical to the one
in Fig. 1. The image was taken at 450 Oe and 4.8 K, and the scale
bar is 0.2 mm long.



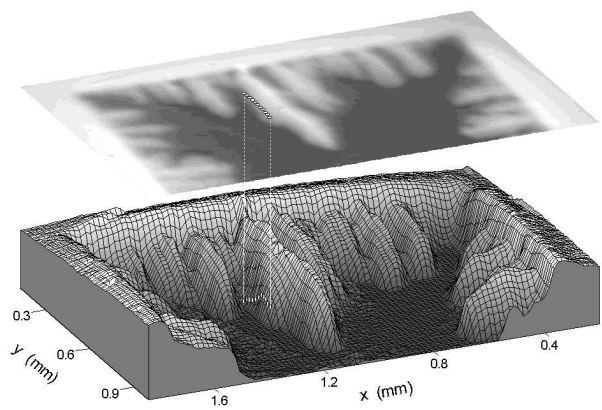
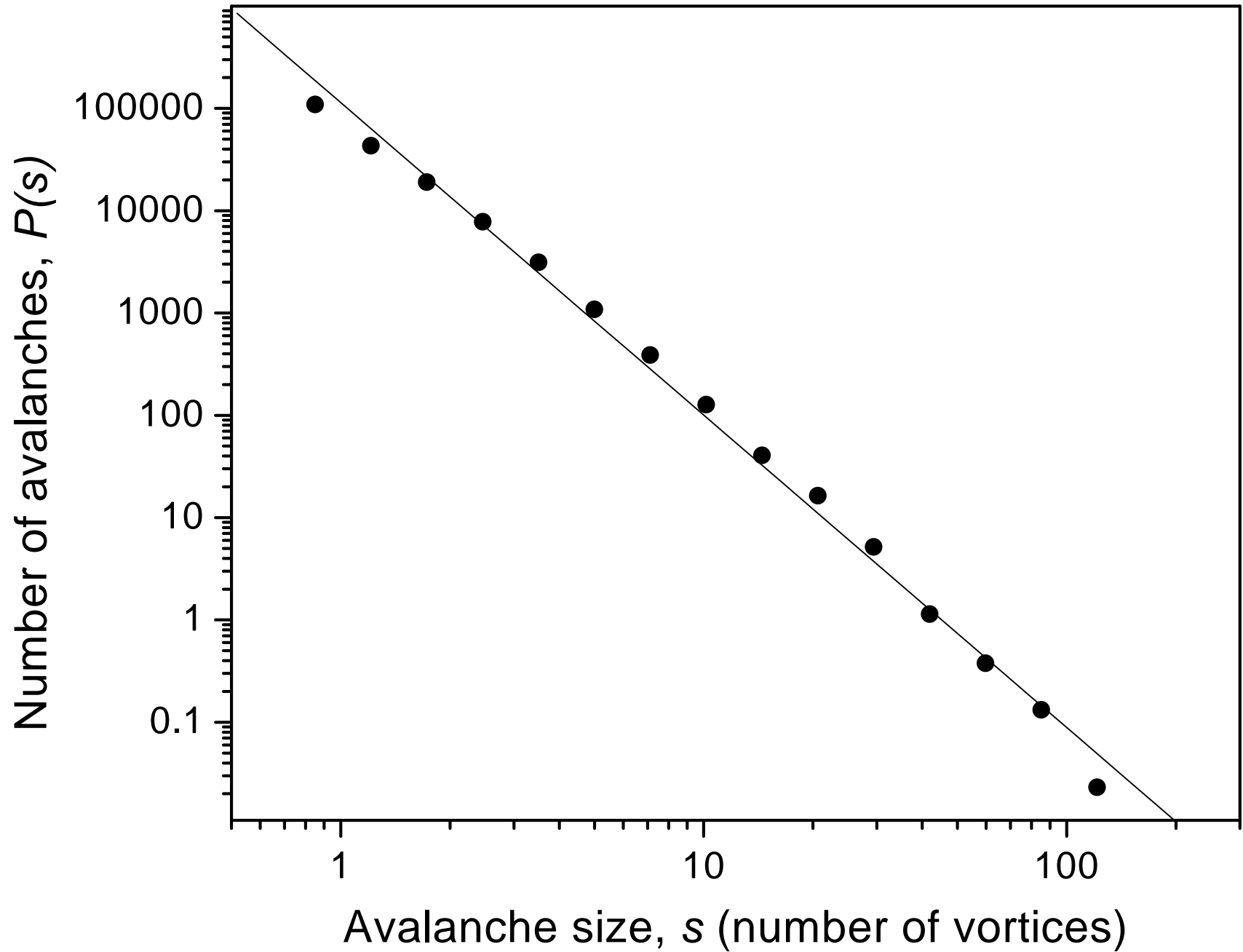
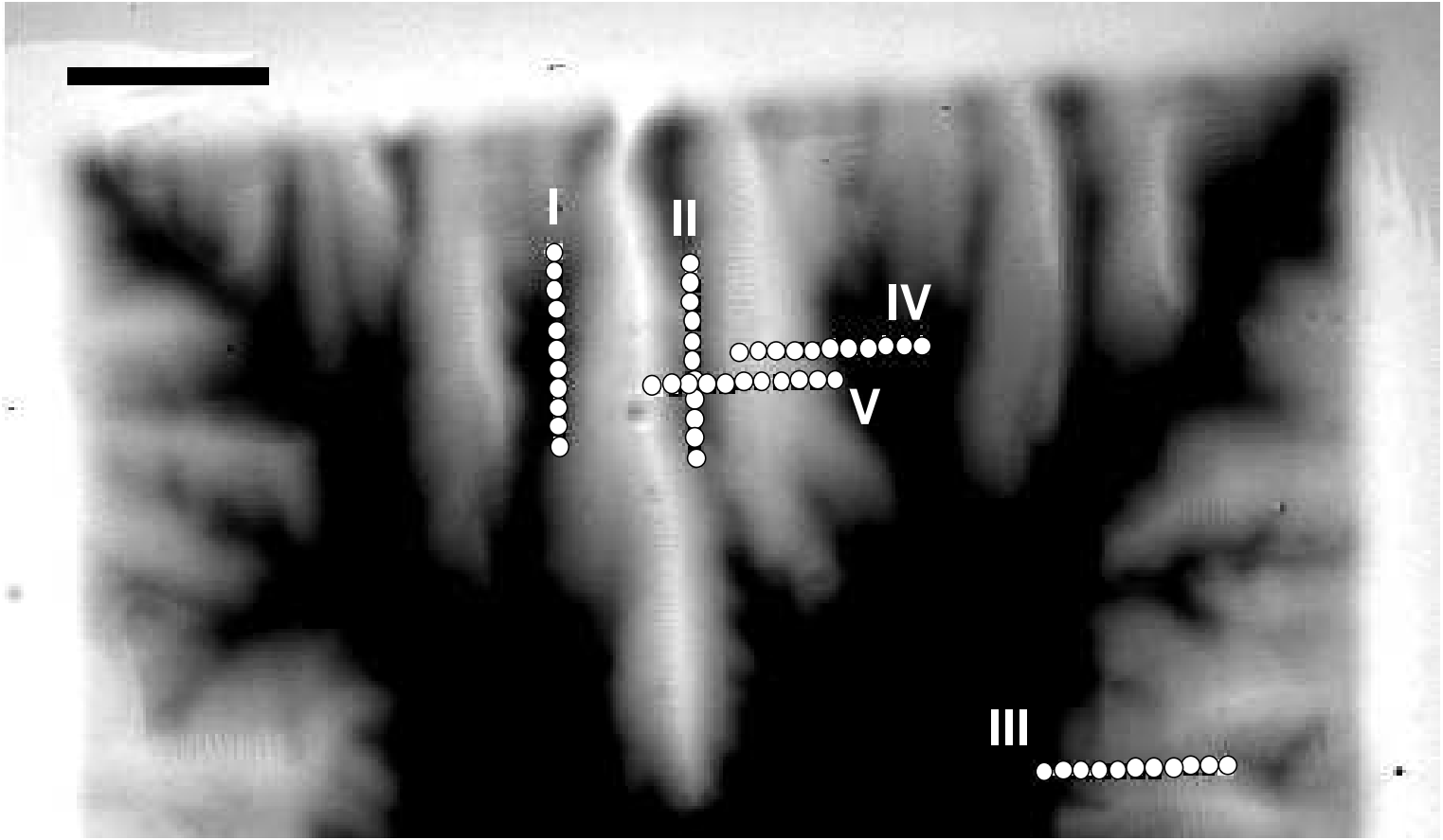
\end{document}